\documentclass[prb,twocolumn,superscriptaddress,nobibnotes]{revtex4}

\usepackage{graphicx}
\usepackage{dcolumn}
\usepackage{bm}
\usepackage{times,mathptmx}
\usepackage{color}
\usepackage{multirow}
\usepackage{enumitem}
\usepackage{chngpage}
\usepackage{amsmath}

\begin{document}


\title{\bf Photon-level tuning of photonic nanocavities}

\author{Mingxiao Li}
\thanks{These two authors contributed equally.}
\affiliation{Department of Electrical and Computer Engineering, University of Rochester, Rochester, NY 14627}
\author{Hanxiao Liang}
\thanks{These two authors contributed equally.}
\affiliation{Department of Electrical and Computer Engineering, University of Rochester, Rochester, NY 14627}
\author{Rui Luo}
\affiliation{Institute of Optics, University of Rochester, Rochester, NY 14627}
\author{Yang He}
\affiliation{Department of Electrical and Computer Engineering, University of Rochester, Rochester, NY 14627}
\author{Jingwei Ling}
\affiliation{Institute of Optics, University of Rochester, Rochester, NY 14627}
\author{Qiang Lin}
\email{qiang.lin@rochester.edu}
\affiliation{Department of Electrical and Computer Engineering, University of Rochester, Rochester, NY 14627}
\affiliation{Institute of Optics, University of Rochester, Rochester, NY 14627}



\begin{abstract}
Energy-efficient optical control of photonic device properties is crucial for diverse photonic signal processing. Here we demonstrate extremely efficient optical tuning of photonic nanocavities, with only photon-level optical energy. With a lithium niobate photonic crystal nanocavity with an optical Q up to 1.41 million and an effective mode volume down to 0.78$(\lambda/n)^3$, we are able to achieve a resonance tuning rate of about 88.4~MHz/photon (0.67~GHz/aJ), which allows us to tune across the whole cavity resonance with only about 2.4 photons on average inside the cavity. Such a photon-level resonance tuning is of great potential for energy-efficient optical switching, wavelength routing, and reconfiguration of photonic devices/circuits that are indispensable for future photonic interconnect.

\end{abstract}

\maketitle

Controlling photonic functionalities by purely optical means is a long-term goal of nonlinear optics, which has been pursued for decades \cite{Cotter99, Agrawal08, Willner14}. Unfortunately, natural optical media generally exhibit fairly weak optical nonlinearities \cite{Boyd08}, leading to significant optical powers required for nonlinear optical interactions. A typical approach to enhance the nonlinear optical effects is to utilize micro-/nanoscopic photonic structures to introduce strong confinement of optical waves \cite{Vahala03, Lukin14}, which effectively increases the optical intensity and energy density to support nonlinear optical interactions. Photonic crystal nanocavities are particularly suitable for this purpose \cite{Soljacic04}, which have been applied on a variety of device platforms for optically tuning/switching the cavity properties \cite{Noda07, Vuckovic07, Vos07, Benisty08, Wong09, Notomi10, Rue10, Gong12, Notomi12, Baba13, Coleman16}.  However, a majority of devices developed to date still require a substantial number of photons, unless a certain quantum emitter (atom, quantum dot, defect, etc.) is incorporated into the cavity to enhance resonant dipole interactions \cite{Lukin14} which unfortunately require sophisticated operation conditions such as high vacuum and/or cryogenic environment, challenging for implementation in a practical environment. To date, it remains a significant challenge to realize device control at room temperature and ambient environment with only photon-level optical energy, which represents the ultimate efficiency of nonlinear optics.

In this paper, we report extremely efficient seamless wavelength tuning of photonic nanocavities, with an optical energy of only a couple photons, via a lithium niobate (LN) photonic crystal nanocavity that exhibits an optical Q as high as 1.41 million, the highest value reported to date for photonic crystal nanocavities made on monolithic LN platform \cite{Bernal12, Diziain13, Wang14, Pertsch14, Liang17, Li18}, to the best of our knowledge. The device simultaneously exhibits an effective mode volume as small as $0.78(\lambda/n)^3$ (where $\lambda$ is the optical wavelength in vacuum and $n$ is the refractive index of LN). The high optical Q and the tiny effective mode volume enable us to achieve extremely efficient all-optical wavelength tuning of the device, with a resonance tuning rate as high as 88.4~MHz/photon (corresponding to 0.67~GHz/aJ), while nearly 100\% preservation of the resonance quality. Consequently, we are able to tune across the full linewidth of the cavity resonance with an optical energy of 0.32 aJ that is equivalent to only about 2.4 telecom-band photons on average inside the cavity.
\begin{figure*}[t!]
	\centering\includegraphics[width=2.0\columnwidth]{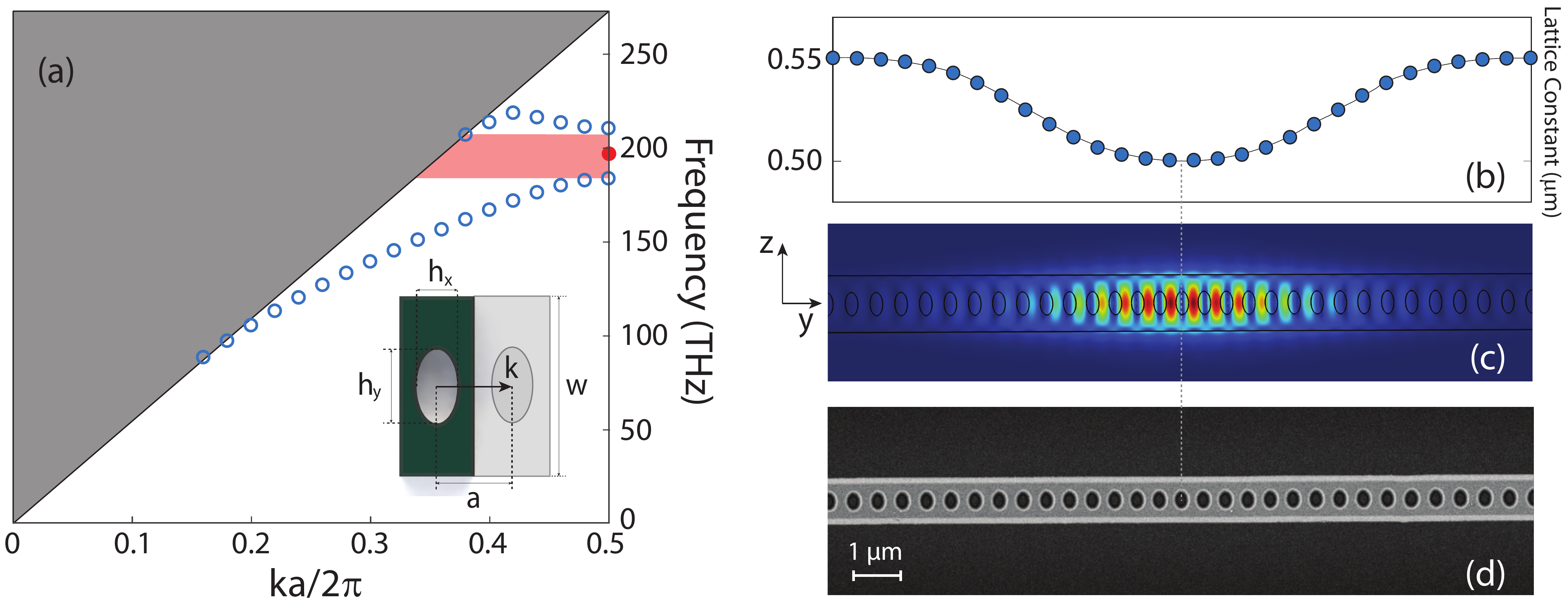}
	\caption{\label{Fig1} Design of LN photonic crystal nanocavity. {\bf (a)} Dispersion property of the designed LN photonic crystal nanobeam, for the TE-like polarization. The inset shows the structure of the unit cell (top view). The red point indicates the frequency of the fundamental TE-like point defect mode and the colored pink region indicates the photonic bandgap. {\bf (b)} Lattice constant as a function of position, which is optimized for high radiation-limited optical Q. {\bf (c)} The optical mode field profile of the fundamental TE-like cavity mode, simulated by the finite element method. The left inset shows the orientation of the LN crystal where the optical axis is along the $z$ direction. {\bf (e)} Scanning electron microscopic image of a fabricated LN photonic crystal nanobeam.}
\end{figure*}

To achieve high optical quality, we employ a photonic crystal nanobeam structure that is formed with patterned elliptical holes, whose unit cell is shown schematically in the inset of Fig.~\ref{Fig1}(a). We swept the width $w$ of the nanobeam, layer thickness $t$, axial lengths $hx, hy$ of the elliptical holes, and lattice constant $a$ to optimize the photonic bandgap. Figure \ref{Fig1}(a) shows the band diagram simulated by the finite element method, where a LN photonic crystal nanobeam with dimensions of $w=1200$~nm, $t=300$~nm, $h_{x}=270$~nm, $h_{y}=490$~nm, and a lattice constant of $a=550$~nm exhibits a photonic bandgap of 26~THz that covers optical frequency from 183 to 209 THz, for the transverse-electric-like (TE-like) polarized guided mode with the electric field dominantly lying in the device plane.

\begin{figure}[b!]
	\includegraphics[width=1.0\columnwidth]{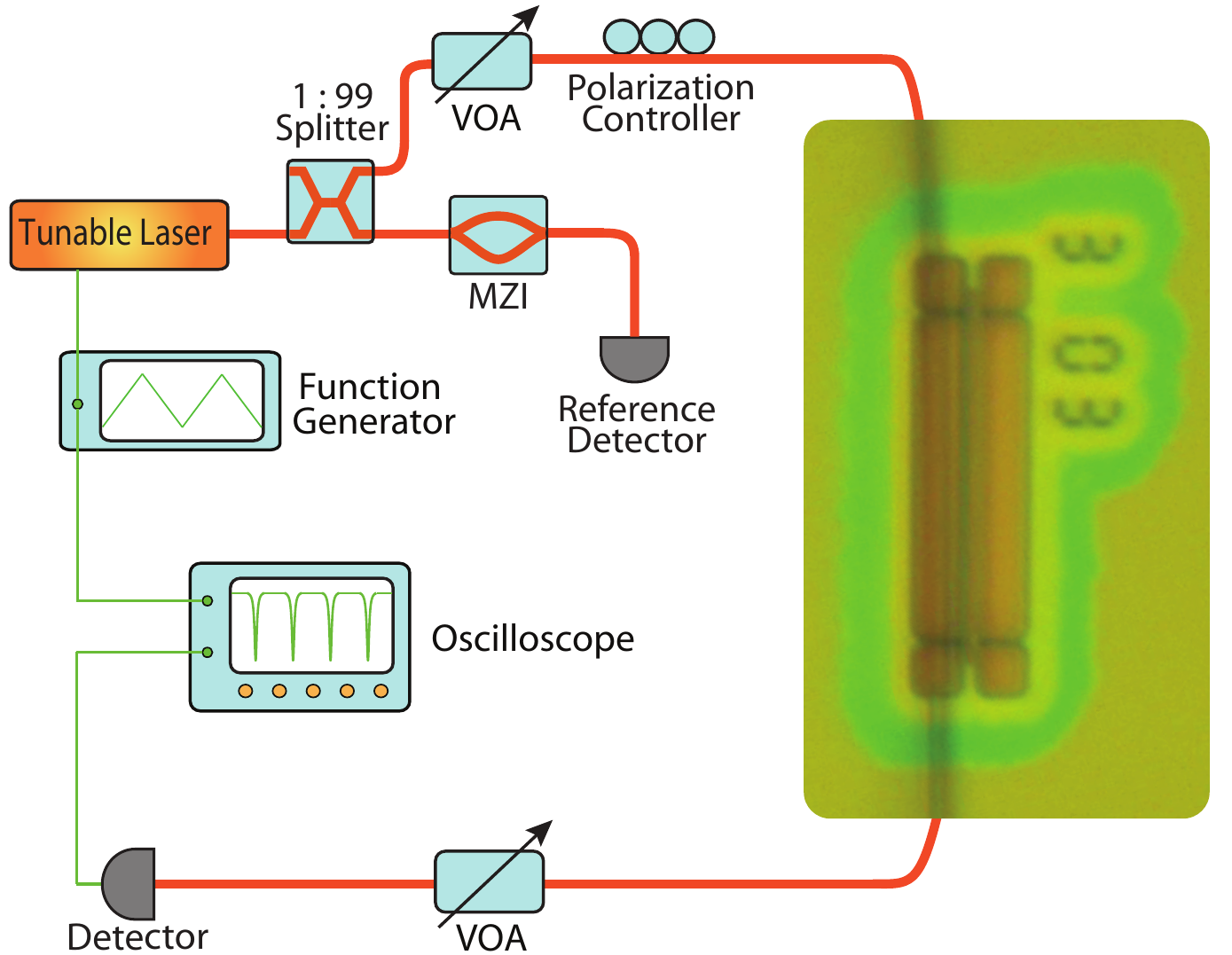}
	\caption{\label{Fig2} Experimental testing setup. Light is coupled into and out of the nanocavity via a tapered optical fiber. VOA: variable optical attenuator; MZI: Mach-Zehnder interferometer.}
\end{figure}
To produce a well-confined point-defect cavity, we gradually varied the lattice constant from 550~nm to 500~nm around the center of the nanobeam. With an optimized pattern of lattice constants shown in Fig.~\ref{Fig1}(b), the photonic crystal nanobeam exhibits a fundamental TE-like cavity mode with a resonance frequency close to the center of the photonic bandgap (red dot in Fig.~\ref{Fig1}(a)) and a spatial mode field profile well localized around the nanobeam center (Fig.~\ref{Fig1}(c)). Detailed simulations by the finite element method show that the cavity mode exhibits a radiation-limited optical Q as high as $1.23\times 10^8$, with effective mode volume as small as 0.78$(\lambda/n)^3$. This level of radiation-limited optical Q is comparable to the best theoretical values reported so far for photonic crystal nanocavities \cite{Benisty08, Noda08, Loncar11}, while LN exhibits a refractive index significantly lower than silicon \cite{Noda08, Loncar11} and gallium arsenide \cite{Benisty08}. As shown in Fig.~\ref{Fig1}(c), the fundamental TE-like cavity mode is a dielectric mode with the optical field dominantly located inside the LN medium, which is crucial for the nonlinear optical process described below.

\begin{figure}[b!]
	\includegraphics[width=1.0\columnwidth]{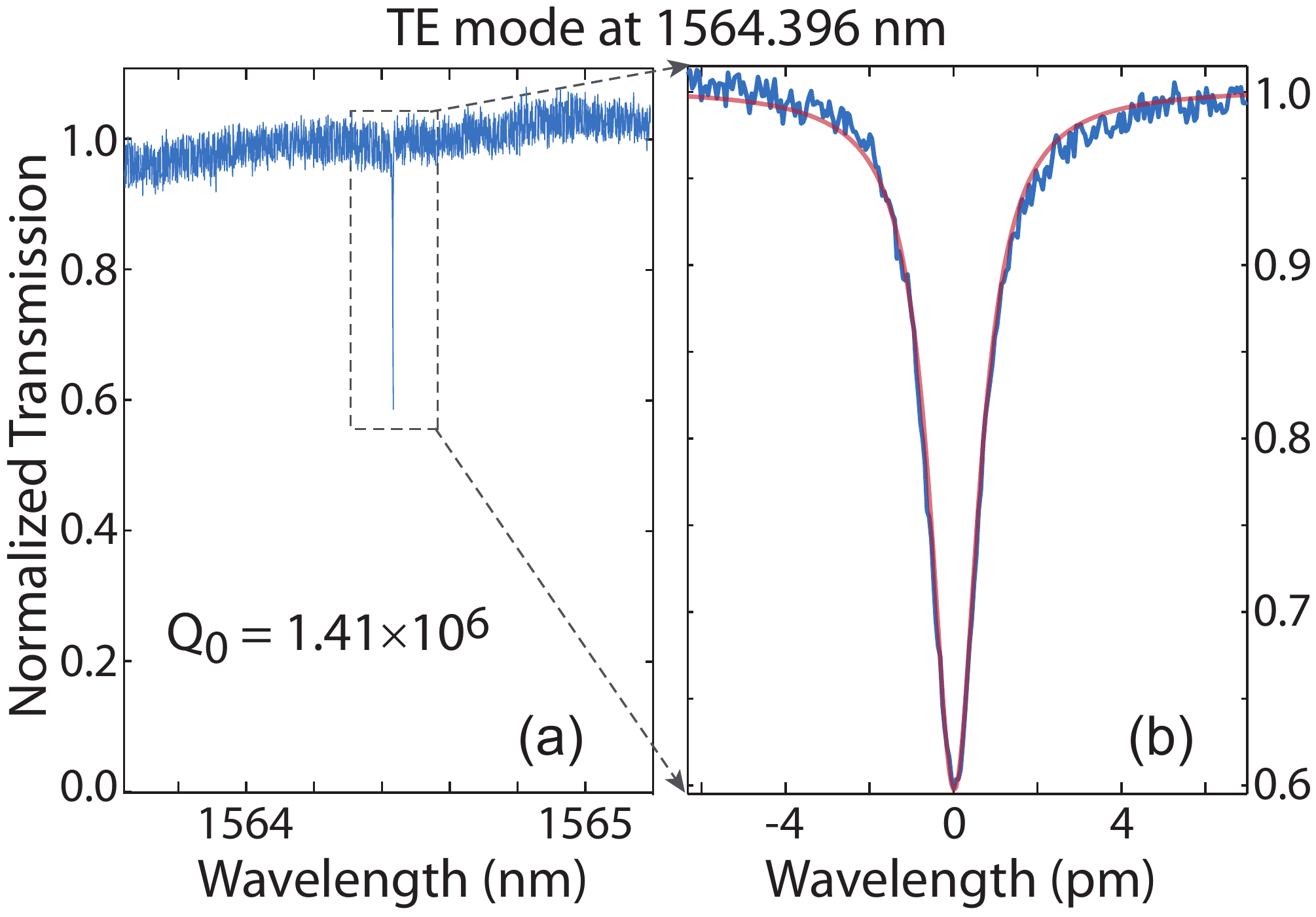}
	\caption{\label{Fig3} Linear optical property of a fabricated LN photonic crystal nanocavity. Laser-scanned transmission spectrum ({\bf a}) and its details ({\bf b}) of the fundamental ${\rm TE}$-like cavity mode, with the experimental data shown in blue and the theoretical fitting shown in red. }
\end{figure}

\begin{figure*}[t!]
	\includegraphics[width=2.0\columnwidth]{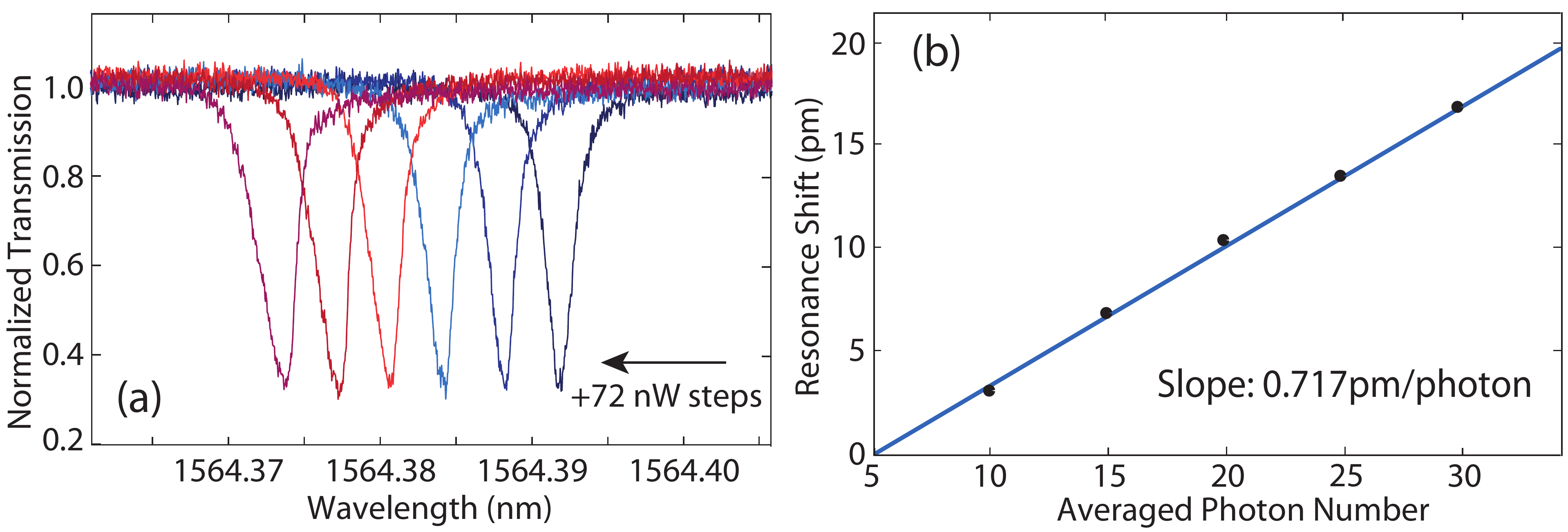}
	\caption{\label{Fig4}  Optical tuning of the cavity resonance. {\bf (a)} Recorded transmission spectra of the cavity at five different input powers from 72 to 432~nW, with a power step of 72~nW. {\bf (b)} Induced wavelength shift of the cavity resonance as a function of averaged number of photons inside the cavity, where the experimental data are shown as black dots and the blue curve is a linear fitting, with a slope of 0.717~pm/photon (corresponding to 88.4~MHz/photon).}
\end{figure*}

The devices were fabricated on a $300$-nm-thick x-cut single-crystalline LN thin film bonded on a 3-${\mu}$m silicon dioxide layer sitting on a silicon substrate. The photonic crystal device structure was patterned with ZEP-$520$A positive resist via electron-beam lithography, which was then transferred to the LN layer with an Ar$^+$ plasma milling process \cite{Loncar14, Li18}. The resist residue was removed by a further O$^+$ plasma etching. Finally, we used diluted hydrofluoric acid to undercut the buried oxide layer to form a suspended photonic crystal structure. Figure \ref{Fig1}(d) shows an example of a typical fabricated device, which shows clearly a well defined and uniform elliptical shape of holes and an accurately patterned lattice structure.

The optical properties of the devices were characterized by the experimental setup shown in Fig.~\ref{Fig2}. A tunable continuous wave laser was launched into a photonic crystal nanocavity via evanescent coupling with a tapered optical fiber which was placed in close proximity to the device and was anchored on the two nanoforks fabricated near the device (Fig.~\ref{Fig2}, inset) for stable operation. The light output from the device is collected by the same tapered fiber and delivered to a detector. The laser wavelength is calibrated by a Mach-Zehnder interferometer. To obtain the optical quality of the device, the laser wavelength is scanned across the cavity resonance and the device transmission is recorded by an oscilloscope.

Figure \ref{Fig3} shows the transmission spectrum of a device. The LN nanocavity shows a fundamental TE-like cavity mode with a resonance around 1564.40~nm, which exhibits an intrinsic optical Q as high as $1.41 \times 10^6$, corresponding to an intrinsic photon lifetime of about 1.2~ns. To the best of our knowledge, this value of optical Q represents the highest optical Q reported to date for photonic crystal nanocavities made on monolithic lithium niobate platform \cite{Bernal12, Diziain13, Wang14, Pertsch14, Liang17, Li18}. It is the second single-crystalline photonic crystal nanocavity other than silicon ones \cite{Noda17} that is able to achieve optical Q above one million.

The extremely high optical quality and the tiny effective mode volume of the LN photonic crystal nanocavity would result in a dramatic enhancement of optical energy density inside the cavity, which would thus support strong nonlinear optical interaction. For lithium niobate, a typical nonlinear optical process is the photorefractive effect \cite{Gunter06, maleki2014lithium} which manifests as a decrease of refractive index whose magnitude depends on the optical energy density inside the device. The photorefractive nonlinearity is known to be probably one of the largest nonlinear optical effects, with a magnitude significantly greater than the optical Kerr effect \cite{Boyd08, Gunter06, Valley94, Atwater10}. Therefore, it can function as an efficient mechanism for all-optical tuning of the cavity properties. However, the photorefractive-induced wavelength tuning demonstrated so far \cite{Maleki06, Breunig16, Liang17} still requires considerable optical energy. The high optical Q and the small mode volume of our device is thus of great potential to push the energy efficiency down to few-photon level, as we will show below.

To show this capability, a simple approach would be to lock the laser wavelength to the cavity resonance and monitor the resonance shift when the laser power changes. Intriguingly, this is not possible in our device since the nonlinear effect is enhanced so strongly that the cavity blockades the laser input whenever we try to lock the laser wavelength to the cavity resonance. As such, to characterize the magnitude of induced resonance tuning, we instead continuously scanned the laser wavelength back and forth across the cavity resonance in a periodic fashion (with a laser wavelength scanning in a triangular fashion, see Fig.~\ref{Fig2}), and monitored the change of the transmission spectrum when the laser power increases.

To do so, we positioned the coupling tapered fiber to maintain a coupling depth of $65$\% for the cavity (Fig.~\ref{Fig4}(a), black curve), which accordingly results in a loaded optical Q around $8.9 \times 10^5$ (corresponding to a linewidth of 1.75~pm for the loaded cavity resonance). The laser wavelength was scanned over a range of 150 pm at a scanning rate of $3$ nm/s. To explore the induced resonance tuning, we increased the input optical power from 72 to 432~nW, which results in an averaged optical power from 0.55 to 3.30 nW coupled into the cavity (since the laser scanning range is 150~pm). This range of coupled optical power corresponds to an averaged intra-cavity optical energy from 0.66~aJ to 3.96~aJ, equivalent to 5 to 30 photons inside the cavity.

Figure \ref{Fig4}(a) shows the cavity transmission spectra at different levels of input laser powers. It shows clearly that the cavity transmission spectrum shifts towards blue when the input power increases. The photorefractive effect primarily tunes the cavity resonance in a dispersive fashion without degrading the resonance quality, as both the spectral shape and the coupling depth remain nearly intact until at a high power level where the transmission spectrum is slightly distorted towards red, which is simply the effect of the thermo-optic nonlinearity \cite{Vahala04}. Figure \ref{Fig4}(b) plots the induced resonance shift as a function of the number of the averaged photons inside the cavity. It shows a clear linear dependence. By fitting the experimental data, we obtained a tuning slope of 0.717~pm/photon, corresponding to $\sim$88.4~MHz/photon. Therefore, about 2.4 photons on average inside the cavity are able to tune across the whole cavity resonance, clearly showing the extreme efficiency of tuning the cavity resonance.

In conclusion, we have demonstrated a lithium niobate photonic crystal nanocavity that exhibits an optical Q of 1.41 million, the highest value ever reported for photonic crystal nanoresonators made on lithium niobate platform. With this device, we were able to achieve extremely efficient tuning of cavity resonance, with a slope of 88.4~MHz/photon, or equivalently, 0.67~GHz/aJ, via the cavity-enhanced photorefractive effect. As a result, we were able to tune across the entire cavity resonance with only about 2.4 photons on average inside the cavity. To the best of our knowledge, this is the first device that is able to achieve photon-level resonance tuning without involving embedded quantum emitters for enhancing resonant light-matter interaction. It is important to note that the demonstrated photon-level resonance tuning is purely a classical phenomenon and does not introduce cavity quantum electrodynamics \cite{Lukin14}, since the underlying photorefractive effect exhibits a slow time response much longer than the cavity photon lifetime \cite{Sun17, Jiang17}.

To date, significant efforts have been devoted to developing various all-optical approaches for photonic signal processing \cite{Cotter99, Agrawal08, Willner14}. However, low energy efficiency has become a major challenge for the application of nonlinear optics in photonic signal processing \cite{Tucker11, Miller17}. The extremely energy-efficient resonance tuning demonstrated here is thus of great promise to resolve the energy-efficiency bottleneck, with great potential for energy-efficient optical switching, wavelength routing, and reconfiguration of photonic devices/circuits that are crucial for a future photonic interconnect.

\section*{Funding Information}
National Science Foundation (NSF) (EFMA-1641099, ECCS-1810169, and ECCS-1842691); the Defense Threat Reduction Agency-Joint Science and Technology Office for Chemical and Biological Defense (grant No.~HDTRA11810047).

\section*{Acknowledgments}

This work was performed in part at the Cornell NanoScale Facility, a member of the National Nanotechnology Coordinated Infrastructure (National Science Foundation, ECCS-1542081).

\end{document}